\documentclass[%
aps,%
twocolumn,%preprint,%
preprintnumbers,%
byrevtex,%
floatfix%
]{revtex4}

\usepackage{epsf}
\usepackage{ifthen}
\usepackage[usenames]{color}
\usepackage{amssymb}
\usepackage{amsfonts}
\usepackage{amsmath}
\usepackage[dvips]{graphicx}
\usepackage[dvips, colorlinks=true, bookmarks, pdftitle={An example PDF file from LaTeX}, pdfauthor={Gianaurelio Cuniberti}, pdfsubject={From LaTeX to PDF}, pdfkeywords={PDF, LaTeX, hyperlinks, hyperref}]{hyperref}                       
\def\DNA{{\sc dna}}
\def\DNA{{DNA}}
\def\ii{\textrm{i}\,}
\def\ee{\textrm{e}\,}
\def\etal{\textit{et al.}}
\def\ie{\textit{i.e.}}
\def\Hc{\textrm{H.c.}}
\def\eg{\textit{e.g.}}
\def\oc{\omega_{\textrm{c}}}

\newcommand{\lsb} {\left[}
\newcommand{\rsb} {\right]}

\newcommand{\lcb} {\left\{}
\newcommand{\rcb} {\right\}}
\newcommand{\lab} {\left\langle}
\newcommand{\rab} {\right\rangle}

\begin{document}
\date{April, 1 2005}
\title{%
Quantum transport through a DNA  wire in a dissipative environment}
\author{R.~Guti{\'e}rrez}
\email[Email: ]{rafael.gutierrez@physik.uni-r.de}
\author{S.~Mandal}
\author{G.~Cuniberti}
\affiliation{%
Institute for Theoretical Physics, University of Regensburg, D-93040 Regensburg, Germany}

\begin{abstract}

Electronic transport through \DNA \  wires in the presence of a strong
dissipative environment is investigated. We show that new 
bath-induced electronic states
are formed within the bandgap. These
states show up in the linear conductance spectrum as a temperature dependent
background and lead to a crossover from tunneling to thermal activated behavior 
with increasing temperature. 
Depending on the strength of the electron-bath coupling, 
the conductance at the Fermi level can 
show a weak exponential or even an algebraic length dependence. 
Our results suggest a new environmental-induced transport mechanism. This might be
relevant for the understanding of molecular conduction experiments in liquid
solution, like those recently performed on poly(GC) oligomers in a water buffer
(B.~Xu {\em et al.}, Nano Lett. {\bf 4}, 1105 (2004)). 

\end{abstract}

\pacs{%
05.60.Gg  %{Quantum transport }
87.15.-v, %{Biomolecules: structure and physical properties}
73.63.-b, %{Electronic transport in nanoscale materials and structures}
71.38.-k, %{Polarons and electron-phonon interactions}
72.20.Ee, %{Mobility edges; hopping transport }
72.80.Le, %{Polymers; organic compounds (including organic semiconductors) }
87.14.Gg, %{DNA, RNA}
}

\maketitle

In the emerging field of molecular electronics, \DNA\ oligomers have drawn in
the last decade the attention of both experimentalists and
theoreticians \cite{schuster04}. This has been mainly motivated by DNA exciting
potential applications which include its use as a template in molecular devices
or by exploiting its self-assembling and self-recognition properties
%\cite{keren03,pompe99}. 
\cite{keren03}. 
Alternatively, DNA strands might act as molecular wires either
in periodic conformations as in poly(GC), or by doping  with
metal cations as is the case of M-DNA \cite{felice04}. As a consequence, the
identification of the relevant charge transport channels in \DNA\ systems
becomes a crucial issue. Transport experiments in \DNA\ derivatives are however
quite controversial~\cite{endres04,porath04}. \DNA\ has been characterized as
insulating \cite{storm01}, semiconducting \cite{porath00} or metallic
\cite{yoo01,tao04}. It becomes then apparent that sample
preparation and experimental conditions are more critical than in transport
experiments on other nanoscale systems. Meanwhile, a variety of factors that
appreciably control charge propagation along the double helix have been
theoretically identified: static  \cite{roche03} and dynamical
\cite{soler03}  disorder related to random base pair sequences
and structural fluctuations, respectively, as well as environmental effects
associated with correlated fluctuations of counterions \cite{barnett01} or
with the formation of localised states within the bandgap
[\onlinecite{endres04,gervasio02}].

%*****************Figure1*************************
\begin{figure}[b]
\centerline{
\epsfclipon
\includegraphics[width=.99\linewidth]{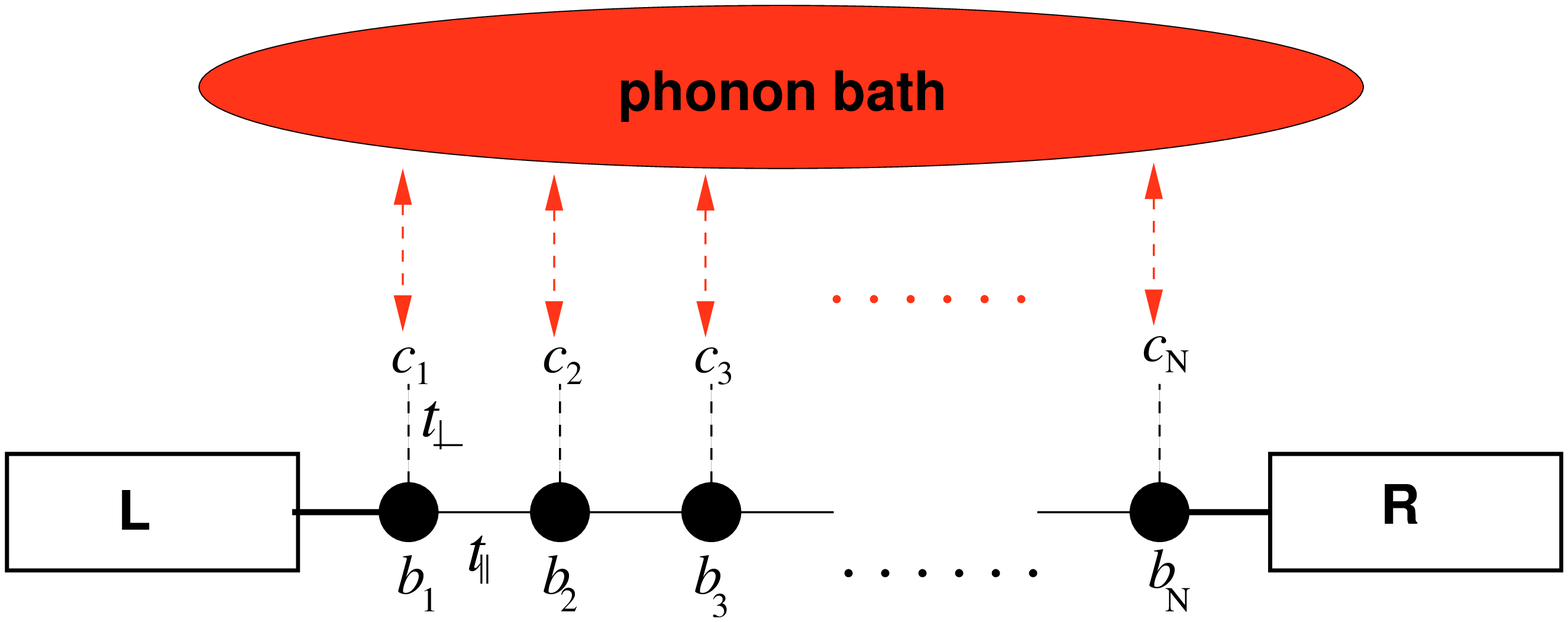}%
}
\centerline{
\epsfclipon
\includegraphics[width=.99\linewidth]{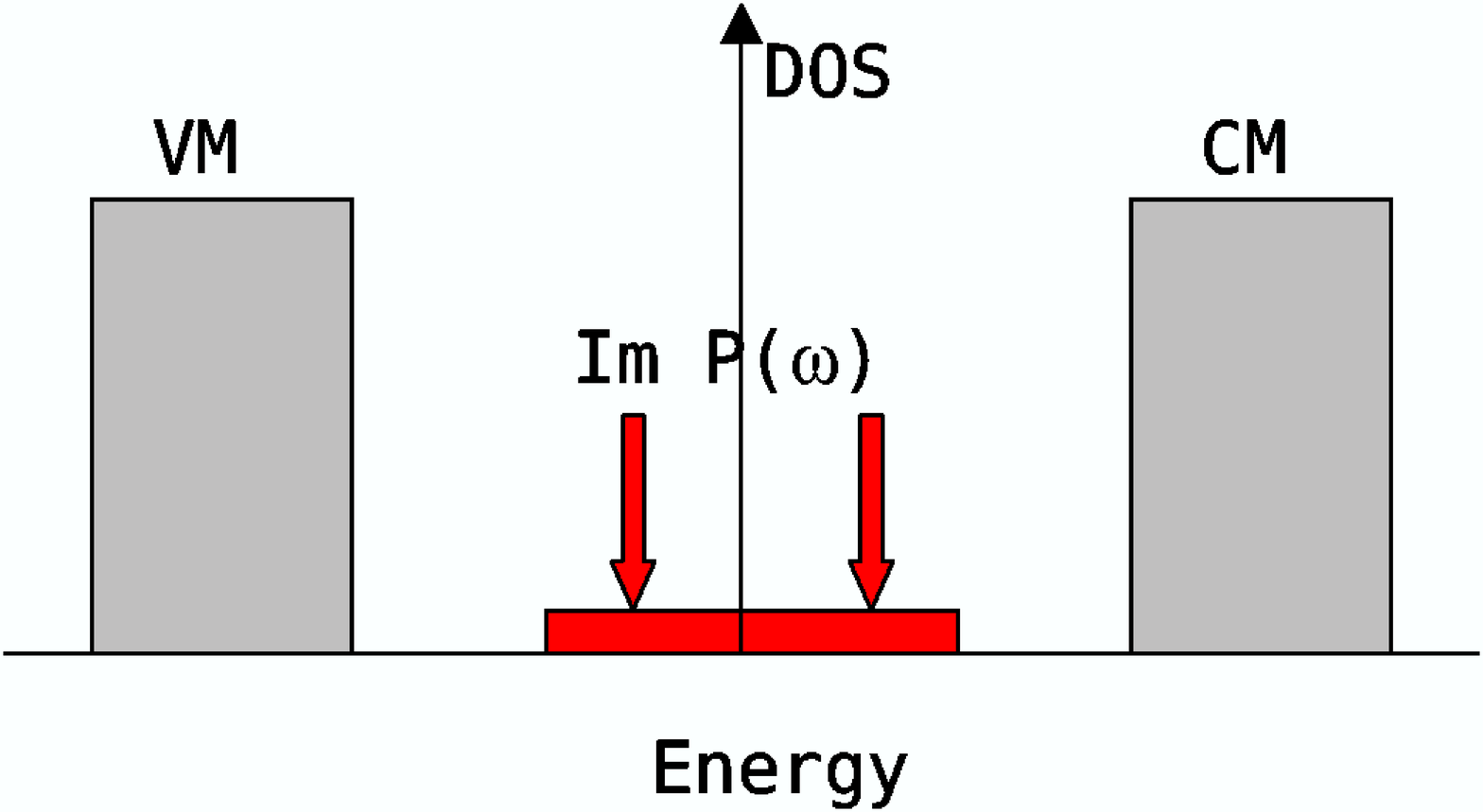}%
\epsfclipoff
}
\caption{\label{fig:fig1}%
Schematic representation of the DNA molecular wire in contact with a phonon bath (upper panel) and
of the corresponding density of states (lower panel). In the absence
of the
phonon bath, valence (VM) and
conduction (CM) manifolds are separated by a gap. Upon
coupling to
the environment, a new set of states emerge within the gap. 
Though strongly damped by the
dissipative coupling, they yield a 
finite density of states and eventually lead to a conductance
enhancement with increasing temperature.
}
\end{figure}
%%%%%%%%%%%%%%%%%%%%%%%%%%%%%%%%%%%%%%%%%%%%%%%%%%%%%%%%%%%%%%%%%%%%%%%%%%%%%%

Recently, Xu \etal~\cite{tao04} have carried out  transport 
experiments on 
poly(GC) oligomers in  aqueous solution. These experiments are remarkable
for different reasons: (i) it was shown that transport characteristics
of {\em single} molecules were probed, 
(ii) the molecules displayed 
ohmic-like  behavior in the low-bias  $I$-$V$
characteristics and (iii) the linear conductance showed an algebraic dependence 
$g\sim N^{-1}$ on the number $N$ of 
base pairs. This latter result suggests the dominance of incoherent 
charge transport processes. Complex band structure calculations \cite{sankey04} for 
{\em dry} poly(GC) oligomers 
predict, on the contrary, a rather strong exponential dependence 
of the conductance on the wire length, 
a  typical result for coherent tunneling through band gaps. 
Hence, Xu \etal~experiments suggest the 
potential role of the  environment
in modifying  the DC conductivity of DNA.

 In the light of these results, we explore in this Letter the
possibility that a strong  perturbation  of the electronic system by a
dissipative environment may lead to a modification of 
the low-energy electronic
structure of the molecular  wire. As a result, the linear transport 
properties may be qualitatively different when comparing with the  ``dry''
wire.

Our description  assumes that only the frontier orbitals of the
poly(GC)
stack are relevant for charge transport, a reasonable approximation at low
bias. Frontier orbitals are  the
highest-occupied (HOMO) and the lowest-unoccupied (LUMO) molecular states. They
both have $\pi$ character and are derived from linear combinations of the $p_z$
orbitals of  individual base pairs. The $\pi$  orbital
stack can be thus represented in a localised orbital picture. 
As shown by first principle calculations \cite{artacho03}, 
the HOMO charge density is 
mainly concentrated  on the guanine bases and the LUMO charge density  on
 the citosyne bases. Hence, within a {\em minimal model},  we will focus 
 \eg~only on the $\pi$-stack along the 
 guanine strand (HOMO) for hole transport and 
 consider the opening of the HOMO-LUMO gap as a perturbation of this
 $\pi$-stack, induced by the  complementary strand and,
 eventually by the backbone subsystem.~\cite{gio02} 
  The  environment is conventionally described by a harmonic phonon
bath.

We address the temperature and length dependence of the conductance 
in the strong coupling limit to the bath
degrees of freedom. Our results can be summarized as follows. 
First, {\it bath-induced} states
appear in the gap region, see Fig.~1 for illustration.
They are however  washed out due to the strong effect of
the environment, so that they do not manifest as well-defined resonances in the
transmission spectrum. Nevertheless, they induce a  temperature
dependent density of states within the gap. 
This leads to a transition from a tunneling regime at low temperatures, 
with a zero current gap, to an  activated regime at higher temperatures, 
with a finite current at low voltages. 
Second, a 
weak exponential or even  algebraic length dependence 
together with   an
Arrhenius-like behavior of the transmission at the Fermi energy are found,
reflecting the strong contribution of incoherent processes.

We describe the system consisting of a poly(GC) wire containing 
$N$ base pairs, contacted to 
left and right electrodes (${\cal H}_{\textrm{leads}}$), 
and in interaction with a phonon
bath (${\cal H}_{\textrm{B}}$) by the following Hamilton operator, see Fig.~1:
\begin{eqnarray}
{\cal H} &=& {\cal H}_\textrm{C} + {\cal H}_{\textrm{C-c}} +{\cal
H}_{\textrm{leads}}+{\cal H}_{\textrm{B}}, \\ {\cal H}_\textrm{C} &=&
\epsilon_b\sum_{j} b^{\dagger}_{j}b_{j} -t_{||} \sum_{j} \left (
b^{\dagger}_{j}b_{j+1} + \Hc \right ), \nonumber \\ 
{\cal H}_{\textrm{C-c}} &=& \epsilon \sum_{j} c^{\dagger}_{j}c_{j} -
t_{\perp}\sum_{j} \left ( b^{\dagger}_{j}c_{j} + \Hc \right ) , \nonumber \\ 
{\cal H}_{\textrm{leads}} &=& \sum_{{\bf k}\in \textrm{L,R}, \sigma} \epsilon_{{\bf
k}\sigma} d^{\dagger}_{{\bf k}\sigma}d_{{\bf k}\sigma} +\sum_{{\bf k}\in
\textrm{L}, \sigma} \left ( V_{{\bf k},1} \, d^{\dagger}_{{\bf k}\sigma} \,
b_{1} + \Hc \right ), \nonumber \\ 
&+&\sum_{{\bf k}\in \textrm{R}, \sigma}
\left ( V_{{\bf k},N} \, d^{\dagger}_{{\bf k}\sigma} \, b_{N} + \Hc \right ),
\nonumber \\ 
{\cal H}_{\textrm{B}} &=& \sum_{\alpha} \Omega_{\alpha}
B^{\dagger}_{\alpha} B_{\alpha} + \sum_{\alpha,j} \lambda_{\alpha}
c^{\dagger}_{j}c_{j} \left (B_{\alpha}+B^{\dagger}_{\alpha} \right ). \nonumber
\label{eq:eq1}
\end{eqnarray}

In the previous expression,${\cal H}_\textrm{C}$  describes 
 the HOMO $\pi$-stack as a
one-dimensional tight-binding chain  with on-site
energies $\epsilon_b$ and intra-strand 
 hopping  $t_{||}$. 
The perturbation arising from the complementary
strand is described via the side-chain Hamiltonian ${\cal H}_{\textrm{C-c}}$. 
The inter-chain hopping $t_{\perp}$ is assumed to be a small 
parameter, according to recent numerical estimates.\cite{anantram05} As
a result, the probability of  inter-chain charge hopping  will be
very small and  we may thus neglect in a first approximation 
 charge propagation
on the side chain, \ie~charge transport occurs only on pathways along 
the central chain.

In Eq.~(1), the onsite energies
$\epsilon_b$ and $\epsilon$ will be set equal to zero for a homogeneous wire. When 
simulating the introduction of  A$-$T bases on a finite segment of the molecular wire, 
we set $\epsilon_b\ne 0$. 
The set of bath frequencies {$\Omega_{\alpha}$} and corresponding coupling
constants {$\lambda_{\alpha}$}, $\alpha=1,\dots , M$, does not need to be further
specified. By performing the thermodynamic limit ($M\to\infty$) later on, the bath can be described 
by a spectral density
$J(\omega)=\sum_{\alpha} \lambda^{2}_{\alpha} \delta(\omega-\Omega_{\alpha}) =
J_0 ( {\omega}/{\oc})^s \ee^{-\omega/\oc} \Theta(\omega)$, where $\oc$ is a
cut-off frequency and $\Theta(\omega)$ is the Heaviside function.~\cite{weiss_book}
 In what follows, we consider only the case 
$s=1$  which corresponds to an ohmic bath. The 
form of the spectral density will of course depend on the specific type 
of environment. In the
case of an aqueous environment, a Debye model for dielectric relaxation might
 seem more
appropriate.~\cite{milena97} 
However, its low-frequency
behavior is similar to that of the ohmic bath; hence, we can safely  approximate
it by the above spectral density. We do not expect that the exponential cut-off
at high frequencies
will have a very dramatic effect on our results.

For $\lambda_{\alpha}=0 \; \forall \alpha$, the model shows a temperature
independent gap in the electronic spectrum, the gap being basically proportional to
$t_{\perp}$. Valence and conduction manifolds, involving $N$ states
each, are symmetric w.r.t.~the Fermi level which is the zero of energy
(particle-hole symmetry). For nonzero coupling to the bath a more involved
behavior may be expected depending on the electron-phonon coupling strength. The
interaction with the bath degrees of freedom can be eliminated by performing a 
unitary transformation~\cite{weiss_book}. As a result the transversal coupling
Hamiltonian ${\cal H}_{{\rm C-c}}$ will be renormalized by the bath
operators~\cite{gmc04}.

 Since we are mainly interested in the temperature and length 
dependence of the linear conductance, we address only the low-bias regime, thus
neglecting  nonequilibrium effects as well as inelastic tunneling which become more 
relevant at large applied voltages. As a result, the current can  still be
written as~\cite{imry04} $I(V)=( {2e }/ {h}) \int \mathrm{d}E \,
(f(E-{eV}/{2})-f(E+{eV}/{2}))\, t(E)$. Note that the function 
$t(E)=4{\textrm{Tr}} \lcb {\rm Im}\mathbf{\Sigma}_\textrm{L} \, \mathbf{G}\,
{\rm Im} \mathbf{\Sigma}_\textrm{R} \, \mathbf{G}^\dagger \rcb$ contains
contributions from the electron-bath interaction via the wire 
Green function  $\mathbf{G}$. It hence describes processes where the 
incoming and outgoing electron energies are equal, though  
 virtual phonon emission and absorption  
is allowed and included to all orders in $\mathbf{G}$. 
We denote in what follows $t(E)$ as a transmission,  though it
is not directly  related to the usual Landauer transmission function 
due to the presence of the dissipative bath in the wire propagator.

 The Green function
$\mathbf{G}$ can be calculated using equation of motion techniques. 
One finds to lowest order in $t_{\perp}$:~\cite{gmc04}
\begin{eqnarray}
{\bf G}^{-1}(E)&=& E{\bf 1}-{\cal H}_\textrm{C}-\mathbf{\Sigma}(E) -
t^2_{\perp} {\bf P}(E), \\ \label{eq:eq2} P_{\ell j}(E)&=&  \delta_{\ell
j}\int_{0}^{\infty} \mathrm{d}t \, \ee^{{\ii}(E+{\ii}0^{+})t}\,
G^{0}_{\rm c}(t)
\, \ee^{-\Phi(t)}, \nonumber
\end{eqnarray}
where 
$\mathbf{\Sigma}(E)=\mathbf{\Sigma}_{\textrm{L}}(E)+\mathbf{\Sigma}_{\textrm{R}}(E)$,
$\ee^{-\Phi(t)}=
\lab{\cal X}(t) {\cal X}^{\dagger}(0)\rab_{\textrm{B}}$ is a dynamical bath 
correlation function and ${\cal X}=\exp{\lsb \sum_{\alpha}
(\lambda_{\alpha}/\Omega_{\alpha})(B_{\alpha}-B^{\dagger}_{\alpha})\rsb}$. 
The electrode selfenergies $\mathbf{\Sigma}_\textrm{L/R}$ are calculated
in the wide-band limit, ${\Sigma}_{\textrm{L},\ell
j}(E)=-{\ii}\Gamma_{\textrm{L}}\delta_{1 \ell}\delta_{1j}$ and
${\Sigma}_{\textrm{R},\ell j}(E)=-{\ii}\Gamma_{\textrm{R}}\delta_{N
\ell}\delta_{Nj}$, \ie~ignoring their energy dependence. Note that the 
function $P(E)$ containing the free electron Green function $G^{0}_{\rm c}(t)$ 
of the side chain and 
the bath correlator $\exp{(-\Phi(t))}$, act as an additional selfenergy for the central chain propagator.

In the weak-coupling regime to the bath, no relevant physical effects were 
found~\cite{gmc04}. In the following, we  discuss the 
strong-coupling limit $J_0/\oc > 1$, where an appreciable modification 
of the electronic spectrum occurs. Recent estimates~\cite{kenzie05}  of the
latter
parameter using the classical Onsager model for  molecule-solvent interactions
suggest that this regime  can be realized in a water environment.

%*****************Figure2*************************
\begin{figure}[t]
\centerline{
\epsfclipon
\includegraphics[width=.95\linewidth]{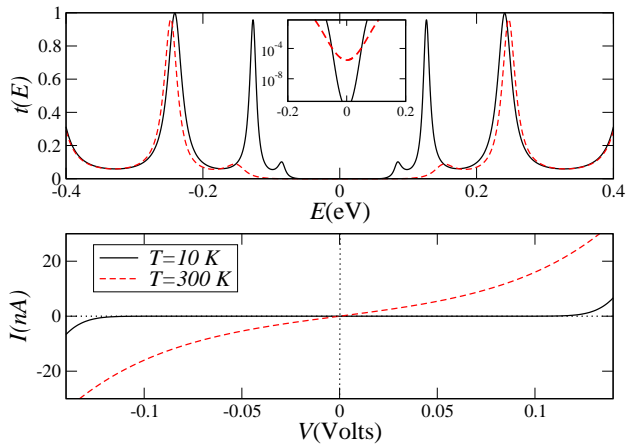}%
\epsfclipoff
}
\caption{\label{fig:fig2}%
Upper panel: The function $t(E)$ for two different temperatures; the inset is a 
log-plot around $E=0$ showing the strong temperature dependence of the 
pseudo-gap.
Lower panel: $I$-$V$ characteristics. Parameters: $N=20$, $J_0/\oc=20, t_{||}=0.6 \, eV,
t_{\perp}/t_{||}=0.2, \Gamma_{\textrm{\textrm{L/R}}}/t_{||}=0.16$.
}
\end{figure}
%%%%%%%%%%%%%%%%%%%%%%%%%%%%%%%%%%%%%%%%%%%%%%%%%%%%%%%%%%%%%%%%%%%%%%%%%%%%%

In Fig.~2 the transmission $t(E)$ and the corresponding current are shown. A strong 
temperature dependent gap
in the electronic transmission spectrum $t(E)$ is found, its magnitude increasing
with temperature. 
The low-voltage $I$-$V$ characteristics evolve from a ``semiconducting'' behavior at low 
temperatures (zero current around $V=0$) to a
``metallic'' behavior (nonzero slope near $V=0$) with increasing temperature. 
The reason is that in the strong dissipative regime  a
{\it pseudo-gap} rather than a gap in the electronic spectrum is induced by the
bath dynamics. An analysis of the real and imaginary parts of $P(E)$, Eq.~(2),
at low energies helps to understand this. 
One can show that (i) ${\rm Re}\, P(E)
\sim E$ for $E\sim 0$ and (ii) ${\rm Im}\, P(E)$ is peaked at $E=0$.
For comparison, in the absence of the bath ${\rm Re }\,P(E)$ would display a
$1/E$ behavior around $E=0$ \cite{gio02}. It follows from (i) that additional
low energy poles of the wire Green function ${\rm \bf G}(E)$ might emerge
symmetrically placed around the Fermi energy, building a third electronic
manifold. These states can show up as resonances in the transmission spectrum inside
the gap, see  Fig.~1, and form a polaronic band. We stress that they are neither present for
$\lambda_{\alpha}=0$ nor in the weak-coupling limit. We would then have 
three electronic manifolds. It turns out, however,
that the non-vanishing ${\rm Im}\, P(E)$ (the ``frictional'' part) has a dramatic influence on these 
states. Since they are located in an energy region where ${\rm Im}\, P(E)$ is
appreciably different from zero, no well-defined resonances manifest in the
low-energy sector of the transmission. Nevertheless, these bath-induced states
{\it do} contribute with a temperature dependent incoherent background and 
eventually lead
to an increase in the density of states near $E=0$ when the temperature grows.
Hence, the current may be enhanced at low voltages with increasing temperature. 
We  
thus interpret the central manifold as an incoherent polaronic band that
supports activated hopping at high temperatures.  This is farther
reflected  in  the temperature  dependence of the transmission at the Fermi
energy, which displays  an Arrhenius-like law, see Fig.~3. We note in passing that a
qualitative similar effect is found in the so called spin-polaron
problem.~\cite{edwards02}

A controversial issue in transport through DNA-based systems is the actual
length dependence of the electron transfer rates or correspondingly, of the
linear conductance [\onlinecite{treadway02,meggers98,kelley99}].
Different functional dependences have been found in charge transfer experiments
ranging from strong exponential behavior related to superexchange mediated electron
transfer [\onlinecite{meggers98}] to algebraic dependences typical of
thermal activated  hopping [\onlinecite{treadway02,kelley99}]. As far as transport
experiments are concerned, Xu \etal~\cite{tao04} reported an algebraic length
dependence of the conductance for poly(GC) oligomers in solution.
Theory has shown that a transition between different regimes may happen as a
function of the wire length [\onlinecite{jortner98}]. We have investigated the
length dependence of $t(E_F)$ and found different scaling laws dependent on 
the strength of the electron-bath coupling. 
For $J_0/\omega_{\rm c}> 1$,
an exponential law for energies close to $E_{\rm F}$ was found, $t(E_{\rm F}) \sim
\exp(-\gamma L)$, see Fig.~3, middel panel. Here, $L=Na_0$, $a_0\sim 3.4\,$\AA $\,$ being the average inter-base
separation. At the first sight, this might be not surprising
since a gap in the spectrum does exist. Indeed, in the absence of the
bath, \ie~with an intrinsic  electronic gap, we get decay
lengths $\gamma_{{\rm coh}}$ of the order of 2 $\, {\rm \AA}^{-1}$. However,
as soon as the interaction with the bath is included, we find values of 
$\gamma$ much smaller than expected for virtual tunneling, ranging from $0.15
\, {\rm \AA}^{-1}$ to $0.4 \, {\rm \AA}^{-1}$. Additionally, $\gamma$ is
strongly dependent on the strength of the electron-bath coupling
 as well as on temperature; $\gamma$ is reduced when $J_0/\oc$ or $k_{\rm
B} T$ increase, see Fig.~3,  
since in both cases the density of states within
the pseudo-gap increases. These results clearly indicate that the bath does
strongly determine the effective decay length in a way which we can quantify by
extracting a $\gamma_{{\rm env}}$ term such that $\gamma=\gamma_{{\rm
coh}}-\gamma_{{\rm env}}$. The first contribution $\gamma_{{\rm coh}}$ is
purely determined by the intrinsic electronic structure of the wire and can be
obtained, \eg~by means of complex band structure approaches \cite{sankey04,fagas04}. 
A $\gamma_{{\rm coh}}$ of the order of $1.5 \, {\rm \AA}^{-1}$ has been 
recently calculated for
poly(GC) \cite{sankey04}, which compares well with our estimated
$\gamma_{{\rm coh}}$. The dependences on $J_0$ and $k_{\rm B}T$ are hence
contained in the bath-induced contribution $\gamma_{{\rm env}}$. 
Remarkably, in the regime $J_0/\omega_{\rm c} \gg 1$   the weak exponential
length dependence goes over onto an algebraic dependence, see Fig.~3, lower panel. 
The introduction of a tunnel barrier induced by the insertion of A$-$T base pairs
in the poly(GC) oligomer \cite{tao04}, can be simulated in the simplest way by a 
shift of the onsite energies along a finite segment of the wire. In this case, the 
exponential dependence is recovered (see the inset of Fig.~3, lower panel).  

%*****************Figure3*************************
\begin{figure}[t]
\centerline{
\includegraphics[width=0.99\linewidth]{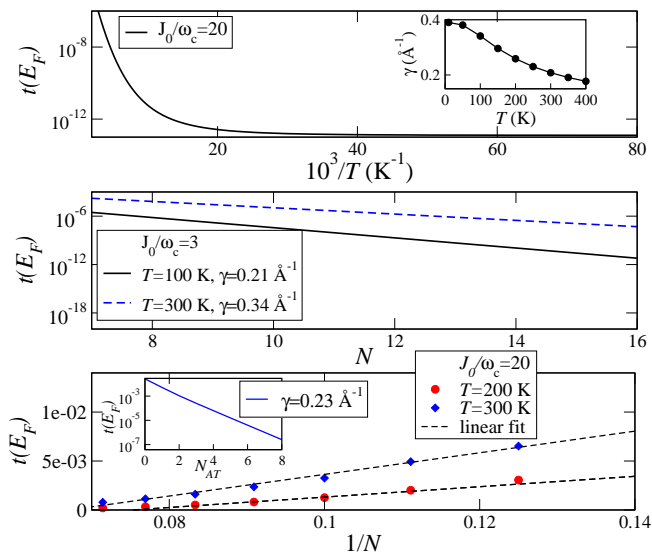}
}
\caption{\label{fig:fig7}%
Upper panel: Arrhenius plot for $t(E_F)$. 
Parameters: $ N=20,\, t_{||}=0.6 \, eV, \, t_{\perp}/t_{||}=0.2,
\,\Gamma_{\textrm{\textrm{L/R}}}/t_{||}=0.16$. Inset: temperature dependence of the
exponential decay length $\gamma$ (see text). 
Middle and lower panels: Length dependence of $t(E_F)$ at different 
temperatures for two different strengths of the electron-bath 
coupling $J_0/\omega_{\rm c}$. The electronic coupling parameters are the same
as in the upper panel.  
The inset in the lower panel shows the effect of introducing  tunnel
 barriers to simulate the insertion
of $N_{\rm AT}$ A-T base pairs in an otherwise homogeneous chain of length $N$
($T$=300 K). The algebraic lenght dependence goes over onto an exponential one.
For this, the onsite energies along a finite 
segment of the wire were shifted by  
$\epsilon_{\rm b}=-1.5 \, eV$, see Eq.~(1).  
The number of unperturbed sites $N_{\rm GC}$ in a 
sequence like $N_{\rm GC}-N_{\rm AT}-N_{\rm GC}$
was kept constant ($N_{\rm GC}=4$) while varying $N_{\rm AT}=1\cdots 8$. 
}
\end{figure}
%%%%%%%%%%%%%%%%%%%%%%%%%%%%%%%%%%%%%%%%%%%%%%%%%%%%%%%%%%%%%%%%%%%%%%%%%%%%%%

In conclusion, we have investigated the influence of a dissipative environment
on charge transport along a molecular wire 
in  a  model  that mimics basic features of the
electronic structure of poly(GC) oligomers. We found a strong modification of the
low-energy electronic structure of the wire in the strong dissipative regime. 
Indeed, a
{\it pseudo-gap} is formed which induces a  temperature dependent
background around the Fermi energy. 
The resulting non-vanishing low-bias current at room temperature as 
well as the algebraic dependence of the conductance on the wire length 
found in
our model 
 suggests that the striking results of 
 Xu \etal~transport experiments~\cite{tao04} may be related to the presence of 
 an aqueous environment. The interplay with other
dynamical degrees
 of freedom  like internal molecular vibrations has, of course, 
 to be further clarified. 
We  note that the
inclusion of randomness in the base pair distribution (as is the case of 
 e.g.~$\lambda$-DNA) does not
qualitatively change the above picture [\onlinecite{gmc04}]. Disorder mainly
washes out the side bands in the transmission without essentially
changing the behavior around the Fermi level.

Finally, we  remark that a close estimation of the physically 
relevant model parameters, 
especially of the electron-bath 
interaction strengths,  requires a detailed analysis of first principle
calculations of DNA oligomers in  solution, eventually including vibrational 
degrees of freedom. 
This goes, however, not only 
beyond the scope
of this investigation, but also beyond the actual capabilities of most {\em ab initio} 
approaches.  

The authors thank P. H\"anggi, J. Keller, M. Grifoni  and M. Hartung for fruitful discussions.
This work has been supported by the Volkswagen foundation and by the EU under
contract IST-2001-38951.

\end{document}